\documentstyle[12pt,aaspp4,psfig,amstex]{article}

\def\arcmin{\hbox{$^\prime$}}
\def\arcsec{\hbox{$^{\prime\prime}$}}

\def\farcm{\hbox{$.\mkern-4mu^\prime$}}
  
\def\lsim{\mathrel{\hbox{\rlap{\lower.55ex \hbox {$\sim$}}\kern-.0em\raise.4ex \hbox{$<$}}}} 
\def\gsim{\mathrel{\hbox{\rlap{\lower.55ex \hbox {$\sim$}}\kern-.0em\raise.4ex \hbox{$>$}}}}

\def\grb{GRB\thinspace{010222}}
\def\smm{SMM\thinspace{J14522+4301}}
\def\ts{\thinspace}

\begin{document}

\title{GRB 010222: A Burst Within a Starburst.}

\author{ 
 D. A. Frail\altaffilmark{1}, 
 F. Bertoldi\altaffilmark{2},
 G. H. Moriarty-Schieven\altaffilmark{3},
 E. Berger\altaffilmark{4},
 P. A. Price\altaffilmark{4},
 J. S. Bloom\altaffilmark{4},
 R. Sari\altaffilmark{5},
 S. R. Kulkarni\altaffilmark{4},
 C. L. Gerardy\altaffilmark{6}, 
 D. E. Reichart\altaffilmark{4},
 S. G. Djorgovski\altaffilmark{4},
 T. J. Galama\altaffilmark{4},
 F. A. Harrison\altaffilmark{4},
 F. Walter\altaffilmark{4},
 D. S. Shepherd\altaffilmark{1},
 J. Halpern\altaffilmark{7}, 
 A. B. Peck\altaffilmark{2}, 
 K. M. Menten\altaffilmark{2},
 S. A. Yost\altaffilmark{4},
 D. W. Fox\altaffilmark{4}
}

\altaffiltext{1}{National Radio Astronomy Observatory, P.~O.~Box O,
  Socorro, NM 87801}

\altaffiltext{2}{Max-Planck-Institut fuer Radioastronomie,
 Auf dem Huegel 69, D-53121 Bonn}

\altaffiltext{3}{National Research Council of Canada, Joint Astronomy
  Centre, 660 N.  A'ohoku Pl., Hilo, HI 96720}

\altaffiltext{4}{California Institute of Technology, Palomar
 Observatory 105-24, Pasadena, CA 91125}

\altaffiltext{5}{California Institute of Technology, Theoretical
 Astrophysics 130-33, Pasadena, CA 91125}

\altaffiltext{6}{Department of Physics and Astronomy, 6127 Wilder
  Laboratory, Dartmouth College, Hanover, NH 03755-3528}

\altaffiltext{7}{Astronomy Department, Columbia University 550 West
  120th St., New York, NY 10027}    

\begin{abstract}
  We present millimeter and submillimeter wavelength observations and
  near-infrared $K$-band imaging toward the bright gamma-ray burst
  \grb. Over seven epochs the flux density of the source was constant
  with an average flux density 3.74$\pm$0.53 mJy at 350 GHz and
  1.05$\pm$0.22 mJy at 250 GHz, giving a spectral index $\alpha=3.78
  \pm 0.25$ (where F$\propto{\nu}^\alpha$). We rule out the
  possibility that this emission originated from the burst or its
  afterglow and we conclude that it is due to a dusty, high redshift
  starburst galaxy (\smm). We argue that the host galaxy of \grb\ is
  the most plausible counterpart of \smm, based in part on the
  centimeter detection of the host at the expected level. The
  optical/NIR properties of the host galaxy of \grb\ suggest that it
  is a blue, sub-L$_*$, similar to other GRB host galaxies. This
  contrasts with the enormous far-infrared luminosity of this galaxy
  based on our submillimeter detection ($L_{\rm Bol} \approx 4 \times
  10^{12}L_\odot$).  We suggest that this GRB host galaxy has a very
  high star formation rate, SFR$\approx$600 M$_\odot$ yr$^{-1}$, most
  of which is unseen at optical wavelengths.
\end{abstract} 



\section{Introduction}\label{sec:intro}

\grb\ was detected on 2001 February 22.31 UT by the gamma-ray burst
monitor on the {\it BeppoSAX} satellite and localized to a error
circle of 2.5\arcmin\ by the Wide Field Camera (Piro
2001a\nocite{piro01a},b\nocite{piro01b}). A bright optical afterglow
candidate was quickly identified by Henden (2001\nocite{hen01}) and
subsequently confirmed by follow-up optical and radio observations
(\cite{mkg+01}, \cite{bf01}). Absorption lines measured by Jha et al.
(2001)\nocite{jpg+01} yielded a lower redshift limit to the GRB of
$z=1.477$, which they argued to be the redshift of the GRB host galaxy
based on the unusually strong equivalent widths of the metallic lines.
Since the burst and its afterglow were bright we initiated GRB
target-of-opportunity programs at several different telescope
facilities (\cite{fkw+00}).

In this paper we report on observations of \grb, which contrary to
expectations, was detected as a constant millimeter and submillimeter
source while the flux density at radio and optical wavelengths, showed
the usual declining behavior expected from the afterglow. After
considering several different origins for the emission, we conclude
that we have detected the host galaxy of the burst, which radiates
primarily in the infrared (IR) and submillimeter, due to the dust and
gas heated by intense star formation. 

\section{Radio Observations}\label{sec:robs}

\noindent
{\it James Clark Maxwell Telescope (JCMT\footnotemark\footnotetext{The
    JCMT is operated by The Joint Astronomy Centre on behalf of the
    Particle Physics and Astronomy Research Council of the UK, the
    Netherlands Organization for Scientific Research, and the National
    Research Council of Canada.}):} A frequency-ordered log of all
observations and flux density measurements are given in Table
\ref{tab:radio}. Observations were made in the 350 GHz and 660 GHz
bands using the Sub-millimeter Common-User Bolometer Array (SCUBA;
\cite{hrg+99}).  In this mode, the secondary mirror is chopped between
the on-source position and a position 60\arcsec\ away in azimuth, in
order to remove sky emission.  The angular resolution of the telescope
is $\sim$14\arcsec\ at 350 GHz and $\sim$8\arcsec\ at 660 GHz.  The
pointing was checked approximately once per hour on a nearby blazar
(JVAS 1419+5423) which is located less than 13$^{\circ}$ from the GRB.
The pointing was generally found to vary by less than 2-3\arcsec.  On
all days where observations were taken, the focus was checked at least
every 2-4 hours, usually on the (local) pointing source.

The data were reduced using the SCUBA User Reduction Facility
(\cite{jl98}).  The raw signals were flat-fielded to account for the
small differences in bolometer response, extinction corrected, and
de-spiked to remove anomalous signals above the 3-sigma level. Short
time-scale sky variations were also removed using pixels around the
edge of the array containing no source emission (\cite{jlh98}).  A
flux calibration factor was then applied to convert to Jy. Flux
calibration factors of 197 $\pm$ 13 Jy/V and 384 $\pm$ 82 Jy/V
were applied to the 350 GHz and 660 GHz data, respectively
(\cite{cou00}).  

The atmospheric opacity at 220 GHz was measured every 10 minutes using
a tipping radiometer mounted near and operated by the Caltech
Submillimeter Observatory.  These values were then extrapolated to 350
GHz and 660 GHz and used to correct for atmospheric extinction
(\cite{awj00}).


\noindent
{\it IRAM 30-meter telescope (IRAM):} Observations were made using the
Max-Planck Millimeter Bolometer (MAMBO; \cite{kgg+98}) array at the
IRAM 30-m telescope on Pico Veleta, Spain.  MAMBO is a 37-element
bolometer array sensitive between 190 and 315 GHz. The larger than
half peak sensitivity range is 210 -- 290 GHz, with an effective
center frequency of $\sim 250$ GHz for steep spectrum sources.  The
beam for the feed horn of each bolometer is matched to the telescope
beam of 10.6\arcsec, and the bolometers are arranged in a hexagonal
pattern with a beam separation of 22\arcsec.  Observations were made
in standard on-off mode, with 2 Hz chopping of the secondary by
32\arcsec.  

The telescope pointing was checked every $\sim 20$ minutes with JVAS
1506+4239, a point source $\sim$2.5 deg from the OT position. The flux
of the pointing source was monitored and appeared stable at 0.15 Jy.
The target pointing accuracy is better than 2\arcsec.  \grb\ was
positioned on the central bolometer of the array, and after each 10
seconds of integration, the telescope was nodded so that the previous
``off'' beam becomes the ``on'' beam. Each scan of sixteen 10 second
subscans lasts 4 minutes, of which 80 sec integration falls on the
source, 80 sec off source, and about 80 sec are lost to move the
telescope and start integration. Gain calibration was performed using
observations of planets, resulting in a flux calibration factor of
12,500 counts per Jansky, which is estimated to be accurate to 15\%.
Opacity corrections were made from frequent skydips. A gain-elevation
correction was applied.

The data were reduced using the MOPSI software package (Zylka
1998)\nocite{zylka98}. The average flux densities found at the target
position along with other observational parameters are given in Table
\ref{tab:radio}.



\noindent
{\it Owens Valley Radio Observatory Interferometer (OVRO):}
Observations were made on two nights at a central frequency of 98.48
GHz with a 1 GHz wide band. On the second night, weather conditions
improved sufficiently to add a second frequency at 220.73 GHz with a 2
GHz wide band. Phase calibration was performed using the source
3C\ts{345}, while observations of Neptune provided the flux density
calibration scale with an estimated uncertainty of $\sim 15$\%.  See
Shepherd et al.~(1998)\nocite{sfkm98}, for further details on
calibration and imaging.  No source was detected on either night.

\section{Optical and Near IR Observations}\label{sec:oobs}

\noindent 
{\it MDM Observatory:} The position of GRB 010222 was observed in the
$K$ band using the TIFKAM infrared imager/spectrograph on the 2.4 m
Hiltner telescope at MDM Observatory on 2001 April 11.  This
instrument uses a $512 \times 1024$ InSb detector, and covers a
$2{\farcm 5}\times 5^{\prime}$ field of view.  A total of 135 dithered
30~sec exposures were sky-subtracted and combined to yield a total
exposure of 67.5~min.  The seeing was
$0.\!^{\prime\prime}8-0.\!^{\prime\prime}9$, and the airmass ranged
from $1.02-1.14$.  Nine standard stars from the list of Persson et al.
(1998)\nocite{pmk+98} were observed throughout the night.  From these
observations we find that the photometric zeropoint accuracy is 0.1
mag due, probably, to the presence of light cirrus.

\begin{figure*}
  \centerline{\hbox{\psfig{figure=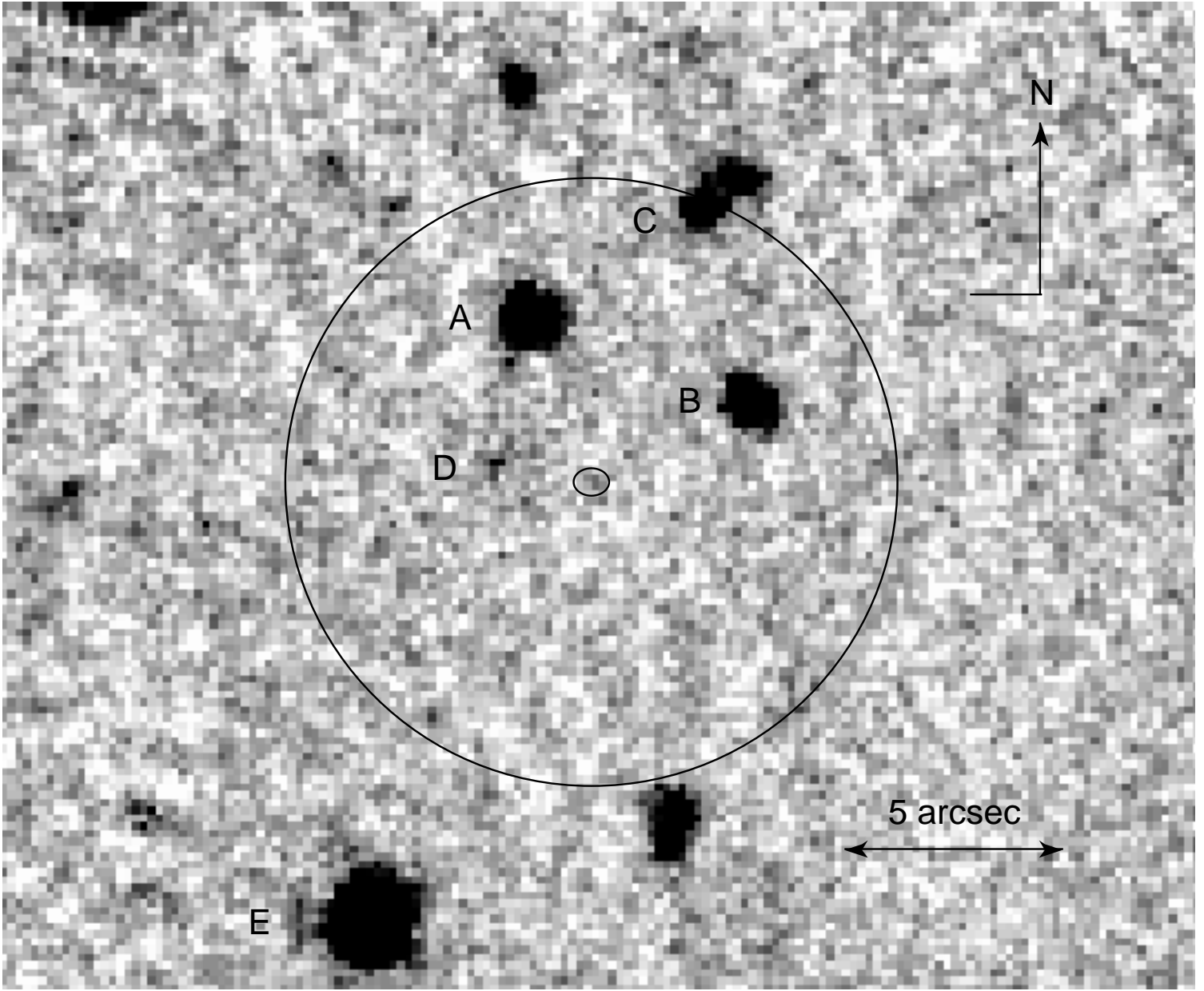,angle=0,width=20cm}}}
\caption[]{K-band Keck II+NIRPSEC image of the field of \grb.  
  The 7-arcsec radius of the SCUBA beam is indicated and the four
  galaxies (A--D) discussed in the text are labeled.  Also marked is
  the $3\sigma$ error ellipse of the optical transient based on a
  relative astrometric solution with the HST+WFPC2 third epoch F814W
  observation of the afterglow.  The host of \grb\ is detected at the
  $3.6\sigma$ level. North is up and east is to the left.
\label{fig:kband}}
\end{figure*}

The magnitudes of several objects in the neighborhood of the GRB host
galaxy were measured. For reference, a magnitude $K= 16.507 \pm 0.033$
was determined for the star ``E'' (Figure \ref{fig:kband}) which lies
10\arcsec\ south and 5.3\arcsec\ east of \grb. Based on this
calibration and the photometry of the faintest visible objects near
the OT, a $4\sigma$ upper limit $K > 19.6$ was derived at the position
of \grb.

\noindent
{\it Keck Observatory:} The location of \grb\ was observed with the
NIRSPEC slit-viewing camera on Keck II\footnote{The W.~M.~Keck
  Observatory is operated by the California Association for Research
  in Astronomy, a scientific partnership among California Institute of
  Technology, the University of California and the National
  Aeronautics and Space Administration.} on 2001 April 27 UT,
approximately 64 days after the GRB.  The observations consisted of a
total of 58 minutes integration over 20 individual dithered frames
with the airmass ranging from 1.3 -- 1.7. The conditions were
photometric with $\sim 0.45$\arcsec\ seeing.  The images were
sky-subtracted and combined using the
IRAF/DIMSUM\footnotemark\footnotetext{{\tt http://iraf.noao.edu/}. The
  DIMSUM package was developed by P. Eisenhardt, M. Dickinson, S. A.
  Standford, and J. Ward with assistance from F. Valdes .} .

The MDM-calibrated magnitude of star E (see above) was used as the
basis of the magnitude zero-point. The host of \grb\ is detected at
the $\sim 3.6\sigma$ level (see Figure \ref{fig:kband}).  For
completeness the four other objects in the $K$-band image that lie
within a 7\arcsec\ radius are indicated in Figure \ref{fig:kband}.
Table \ref{tab:host} contains the results of the $K$-band and F814W
photometry and the (F814W$-K$) colors (see below).  Aperture
photometry was performed on each of these sources with circular
apertures whose radius was chosen to include most of the object flux,
based on a curve of growth.

In order to derive colors for these galaxies matching apertures were
used (with re-centering allowed) to photometer HST+WFPC2 F814W images
of the field. These HST data were taken over four epochs and are part
of a multi-epoch, broadband monitoring of the afterglow of \grb. Full
details of the HST observations can be found in Galama et al., ({\it
  in prep}).

\section{The Origin of the Submillimeter and Millimeter
  Emission}\label{sec:origin}

\begin{figure*}
  \centerline{\hbox{\psfig{figure=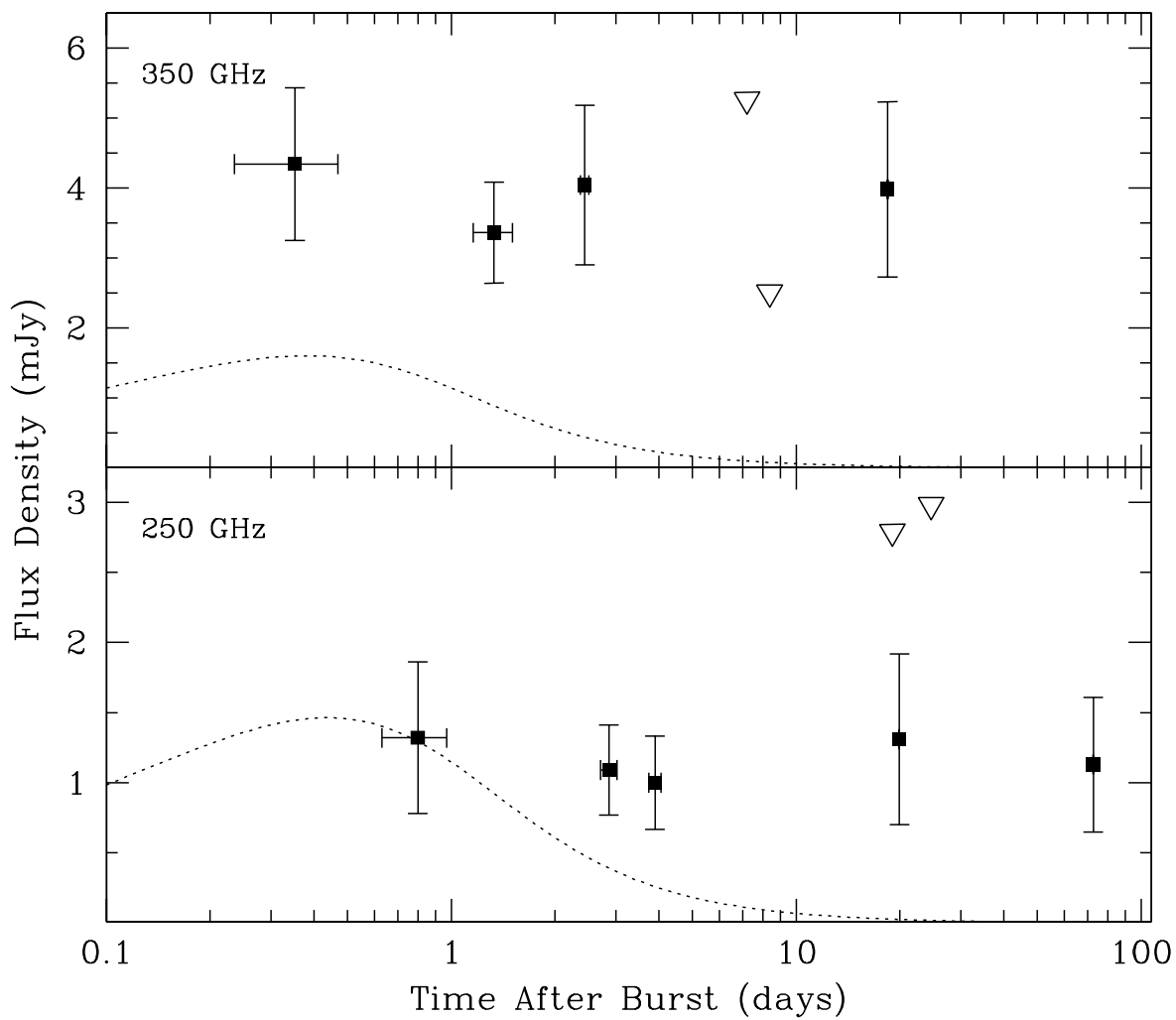,angle=0,width=17cm}}}
\caption[]{Flux density measurements toward \grb\ taken with the 
  SCUBA bolometer array on the James Clerk Maxwell Telescope (top
  panel) and with the Max-Planck Millimeter Bolometer array (MAMBO) on
  the IRAM 30-meter telescope (bottom panel).  Upper limits are shown
  as open triangles and are plotted as the peak flux density at the
  location of the afterglow plus two times the rms noise. The
  observing frequency is shown in the upper left corner of each panel.
  The dotted lines are afterglow light curves derived from a global
  fit to the entire broad-band dataset (see Galama et al., {\it in
    prep}).
\label{fig:one}}
\end{figure*}

Emission was detected towards \grb\ at 250 GHz and 350 GHz (see Table
\ref{tab:radio}). In contrast to the steady decay of the centimeter
and optical emission, the submillimeter flux at 350 GHz remains
remarkably constant from 6 hours after the burst, until 18 days later
(Figure \ref{fig:one}). The noticeable exception to this behavior is
on Day 8, when the measured flux density lies 2.4$\sigma$ below the
average of all measurements.  The data were examined carefully for
errors and inconsistencies in the calibration and data taking, but
apart from some problems with the telescope focus at this time
(R.~Ivison, priv.~comm.), we can find no instrumental origin for the
deviation.  Nevertheless, significant fluctuations in photometric
measurements taken with single dish submillimeter telescopes are not
uncommon. For example, the 350 GHz photometry of the $z=1.44$
ultraluminous starburst galaxy ERO J164502+4626.4 reported by Cimatti
et al.\ (1998)\nocite{cart98} is twice the correct value (Dey et al.\ 
1999)\nocite{dgi+99}.  A search for variability was carried out within
individual days but no statistically significant evidence for any
variations was found. Adding the 350 GHz data from all days together
gives an average flux of 3.13$\pm$0.47 mJy. Excluding the low
measurement gives an average of 3.74$\pm$0.53 mJy.

Independent support that the source is constant comes from the IRAM
250 GHz measurements. Although the flux level is weaker than at 350
GHz (Figure \ref{fig:one}), the average flux density at 250 GHz over
the same time interval (found by averaging all observations while
weighting each scan with the inverse square of its average noise
level) was 1.05$\pm$0.22 mJy. A later measurement in 2001 May 5 gives
a consistent value of 1.13$\pm$0.48 mJy. The 350 GHz to 250 GHz flux
ratio of these average flux density values implies a steep spectral
index $\alpha=3.78 \pm 0.25$ (where F$\propto{\nu}^\alpha$).

The constancy and steep spectral slope of the submillimeter and
millimeter emission towards \grb\ are anomalous compared to the
behavior of past GRBs and when compared to observations of this event
at other wavelengths. In particular, the temporal behavior is
difficult to reconcile with the expectations of the simplest afterglow
models.  This point is illustrated in Figure \ref{fig:one} where we
show the deviation of the 250 GHz and 350 GHz measurements from the
predicted afterglow light curves, as determined by broadband fitting
of the radio, optical and X-ray data (Galama et al., {\it in prep}).

\begin{figure*}
  \centerline{\hbox{\psfig{figure=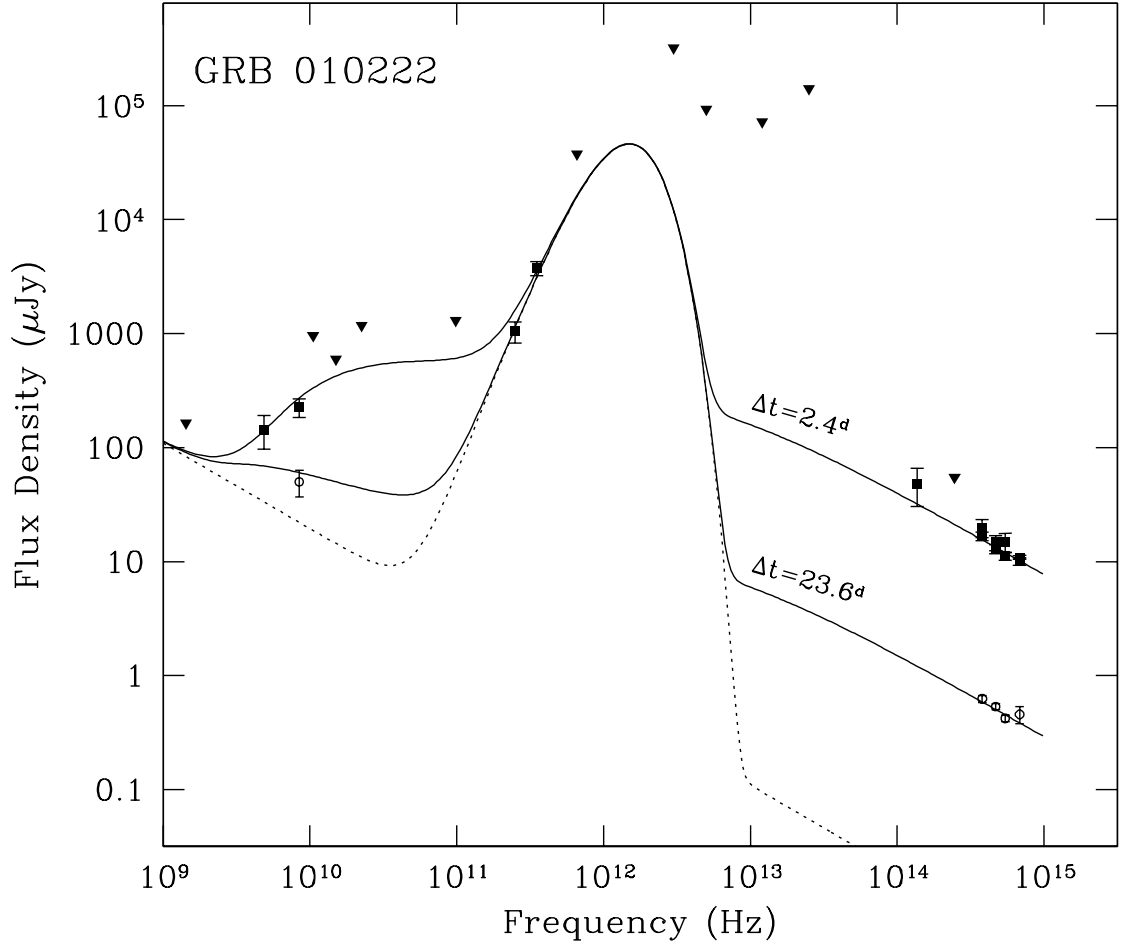,angle=0,width=16cm}}}
\caption[]{Spectral energy distribution (SED) towards \grb. The   
  dashed line is the contribution of the ULIRG \smm\ to the total
  emission. It consists of a modified blackbody of the form
  $F_\nu\propto B_\nu({\rm T})\{ 1-exp[-(\nu/\nu_\circ)^\alpha]\}$
  (where T=40 K, $\alpha=1.5$ and $\nu_\circ=10^{15}$ Hz) plus a
  power-law synchrotron spectrum of the form
  $F_\nu\propto\nu^\alpha_r$ where $\alpha_r=-0.75$ (see text for more
  details). To this starburst SED we have added the SED for the
  afterglow emission from \grb\ (solid lines) at two epochs, 2.4 and
  23.6 days after the burst.  These have been derived from broadband
  modeling of all available radio, optical and X-ray measurements
  taken between 0.15 days to 75 days after the burst (Galama et al.,
  {\it in prep}). The millimeter and submillimeter flux density
  measurements are from Table \ref{tab:radio}, the centimeter radio
  and NIR/optical data are from Galama et al., and the IRAS data were
  obtained from the NASA/IPAC InfraRed Science Archive. To better
  constrain the SED of \smm\ we have used the average flux densities
  at 250 GHz (IRAM) and 350 GHz (JCMT), rather than single epoch
  measurements.
\label{fig:two}}
\end{figure*}

Similar difficulties are posed by the peculiar spectral slope (see
Figure \ref{fig:two}), whose value between 250 GHz and 350 GHz is
$\alpha=3.78 \pm 0.25$. Free-free or synchrotron self-absorption
(\cite{wax97a}, \cite{kp97}) in the millimeter band can lead to steep
slopes with $\alpha=2-2.5$.  However, this explanation fails because
the source was readily detectable at centimeter wavelengths
(\cite{bf01}) where a simple extrapolation of this slope should
suggest otherwise.

Such steep slopes are reminiscent of thermal emission from dust.  At
wavelengths longer than optical, the emissivity of dust grains is
usually modeled as $Q(\lambda)\propto \lambda^{-\beta}$ where $1 \le
\beta \le 2$. For dust which is optically thin in the Rayleigh-Jeans
portion, this results is a spectrum of $F_{\nu} \propto
\nu^{2+\beta}$, consistent with the observed steep slope.  We begin by
considering whether the dust could have been heated (and partially
evaporated) by the burst itself or by the prompt optical/UV flash
generated in the reverse shock of the afterglow (\cite{wd00},
\cite{eb00}, \cite{rei01}, \cite{vb01}).  The subsequent reprocessing
of this radiation, either from dust scattering or by thermal emission,
results in a long-lived infrared emission component (aka dust echo). A
simple energy calculation demonstrates the difficulties with this
interpretation. For our adopted cosmology (H$_0$=65~km
s$^{-1}$~Mpc$^{-1}$, $\Omega_M$=0.3, and $\Lambda_0$=0.7) the
luminosity distance corresponding to the lower redshift limit of
$z=1.477$ (\cite{jpg+01}) is $d_L=3.56\times{10}^{28}$ cm. The
isotropic luminosity at 350 GHz $\propto\nu\times F_\nu$ (defined in
the rest frame) is $L_\nu\simeq 5\times{10}^{44}$ erg s$^{-1}$.
Within 8 days of observation, the total energy emitted directly into
the 350 GHz band is of order $10^{50}$ erg. This dust emission is
expected to be roughly isotropic, so there should be no geometric
correction even if the burst itself was beamed.  More importantly, the
bolometric corrections are expected to be huge.  Calculations by
Waxman \& Draine (2000) and Venemans \& Blain (2001) show that dust is
sublimated by the optical/UV flash at any distance closer than about
$10$ pc, and the surviving dust just outside this distance is heated
to 2000 K, with a spectral peak at a rest-frame wavelength of 1-10
$\mu$m. In this case the bolometric correction to $L_\nu$ at 350 GHz
is more than six orders of magnitude.  Even in the most optimistic
case, the optical/UV flash is only expected to be a small fraction of
the GRB energy (\cite{sp99a}, \cite{wd00}) and therefore it is not
enough to account for the total energy radiated at 350 GHz during the
duration of our observations.  If one hypothetically sets up a
geometrical situation in which the dust is located farther away from
the burst, to allow for lower temperatures and therefore smaller
bolometric correction to the 350 GHz flux, one finds that the emission
will last for years, increasing again the required energy.  We
therefore conclude the GRB is not energetic enough to create the
sub-mm flux by heating its surrounding dust.

We are led to a model in which the millimeter and submillimeter
emission is not dominated by the direct or reprocessed light from the
afterglow itself, but rather originates as a distinct and separate
component.  The simplest hypothesis, consistent with the persistent
nature of the source and its steep spectrum, is that we have detected
a dusty, high redshift galaxy coincident with \grb, designated
as \smm. The population of submillimeter galaxies, of which \smm\ is
an example, produce the bulk of the submillimeter background and are
an important contributor to star formation at high redshifts
(\cite{sib97}), such systems appear to be analogs of local
ultraluminous infrared galaxies (ULIRGs, {\it e.g.}, \cite{sm96}). In
the next two sections we will examine this hypothesis in more detail
and derive the physical properties of galaxy.

\section{Submillimeter and Millimeter Galaxy Properties}\label{sec:properties}

If \grb\ occurred in \smm, then its physical properties can be
derived.  The FIR and submillimeter emission from \smm\ originates
from dust, heated by an interstellar radiation field dominated by
massive stars (\cite{ken98}). In this instance we are well-justified
in representing the thermal dust emission spectrum by a modified
blackbody which takes the form in the Rayleigh-Jeans regime of
$F_\nu\propto \nu^\alpha \times B_\nu(T)$.  A fit of the OVRO, IRAM
and JCMT flux density measurements (and upper limits) results in a
bolometric luminosity of $L_{\rm{Bol}}=4\times{10}^{12}$ L$_\odot$. A
dust temperature of T$_d$=40 K and a dust emissivity index $\beta=1.5$
were assumed (\cite{blain99}). For a continuous starburst, the star
formation rate SFR$\simeq\delta_{\rm MF}$(L$_{\rm{Bol}}/10^{10} {\rm
  L}_\odot)$ M$_\odot$ yr$^{-1}$ (\cite{ocb+01}), where $\delta_{\rm
  MF}$ depends on the stellar mass function. At starburst ages of
10-100 Myr, $\delta_{\rm MF}$ lies between 0.8 and 2, so adopting
$\delta_{\rm MF}$=1.5 we derive SFR$\simeq$600 M$_\odot$ yr$^{-1}$.
Although the uncertainties are at least a factor of two, these values
of SFR and L$_{\rm{Bol}}$ place \smm\ in the class of ULIRGs
(\cite{sib97}).

With what degree of confidence can we assume that \smm\ is the host
galaxy of \grb? From submillimeter source counts of dusty galaxies
(\cite{bki+99}) background source confusion is expected at the flux
level of \smm\ in the 14.5\arcsec\ (FWHM) JCMT beam at 350 GHz
2$\pm$1\% of the time. Given the surfeit of NIR/optical candidates in
Figure \ref{fig:kband}, it is not sufficient to secure the
identification of \grb\ with \smm\ on {\it a posteriori} grounds
alone.  Similar claims have been made for an {\it ISO} satellite
detection of a persistent source towards GRB\ts{970508} at 60 $\mu$m
(\cite{hlm+00}). In our view, independent corroborative evidence is
needed before any such associations can be made.

The problem here is a common one. Because of the large beams used in
their detection, source confusion poses considerable difficulties for
identifying the optical counterparts of submillimeter-selected
galaxies ({\it e.g.,} \cite{sibk00}). In addition to a detection
coincident with \grb, there are four other objects in the $K$-band
image that lie within the JCMT beam (Figure \ref{fig:kband} and Table
\ref{tab:host}). Deep observations ($I>26$ and $K>21$) towards
submillimeter sources have revealed a growing number of highly
obscured sources (EROs) with very red colors ($I-K)\gsim 5$
(\cite{sik+99}, \cite{dgi+99}, \cite{fsi+00}). Object ``B'', 4\arcsec\ 
away from \grb\ is the reddest galaxy in the error circle with
(F814W$-K)\simeq 4.5$, nominally qualifying it as an EROs.  The red
colors of ERO galaxies may be due to an old stellar population or a
highly reddened population of young stars. Their space density,
although not well determined, are comparable to that of SCUBA sources
at these flux levels (\cite{cgh+94}, \cite{mdgs97}, \cite{sik+99}).

However, not all submillimeter galaxies are red objects. A case in
point is SMM\thinspace{J14011+0252} at $z=2.56$, (\cite{isb+00}) which
consists of two components (J1 and J2) neither of which are especially
red ($I-K)\sim 3$. Still others, such as SMM\thinspace{J02399$-$0136}
at $z=2.80$ (\cite{isl+98}), and SMM\thinspace{J02399$-$0134} at
$z=1.06$ (\cite{skb+99}) have blue colors. This diversity of optical
properties is also reflected in samples of local ULIRGs, which have
SEDs that when shifted to high $z$ can be as blue as
optically-selected field galaxies (\cite{tks99}).

The detection of centimeter radiation coincident with submillimeter
sources has proven to be a powerful method for identifying
counterparts (\cite{sio+00}). Likewise, the most secure identification
of an ULIRG with a GRB is that of Berger et al.~(2001)\nocite{bkf01},
where a persistent centimeter source was found spatially coincident
with GRB\ts{980703}. The method relies on the well-known radio/FIR
correlation for star-forming galaxies (\cite{con92}). The thermal dust
emission at 350 GHz due to massive stars can be related to the
synchrotron and thermal Bremsstrahlung emission at 1.4 GHz via a
redshift-dependent spectral index $\alpha^{350}_{1.4}$ (\cite{cy99}).
Using the relation of Carilli \& Yun (2000) we find a value of
$\alpha^{350}_{1.4}=0.69\pm{0.16}$ at $z=1.477$, and predict a
centimeter flux density at 1.4 GHz of 85$^{+115}_{-50}$ $\mu$Jy.
Galama et al.~(2001) have noted a flattening of the 8.46 and 4.86 GHz
light curves at late times. This deviation from model predictions of
the afterglow flux is significant at the 2.5-$\sigma$ and 2.3-$\sigma$
levels, respectively. For a centimeter spectral index of
$\alpha=-0.7$, typical of such galaxies, we find that within the
errors the centimeter emission is consistent with an ULIRG at $z\simeq
1.5$. There are no other radio sources visible in the JCMT beam to an
rms $\sim$6 $\mu$Jy beam$^{-1}$ with a $\sim$1\arcsec\ beam at 8.46
GHz. Continued monitoring will verify whether this is the centimeter
detection of the host galaxy of \grb.

In the absence of suitable optical/NIR color discriminants,
submillimeter sources have been identified by searching for molecular
gas at the candidate redshift ({\it e.g.,} \cite{fis+98}), or by
better constraining the position with a interferometric detection of
the millimeter continuum ({\it e.g.,} \cite{bcm+00}). If \smm\ is an
ULIRG at the same redshift of \grb, then it is estimated from the
observed flux density at 250 GHz (and assuming T$_d$=40 K,
$\beta=1.5$, dust-to-gas ratio=100) a total gas mass of 3$\times
10^{10}$ M$_\odot$ (see Omont et al. 2001\nocite{ocb+01} and
references therein) - an amount that is readily detectable with
current instruments. A future search for redshifted CO(2-1) at
$z=1.477$ would be a promising means to confirm the association.

\section{Optical and NIR Galaxy Properties}\label{sec:onirprop}

The current evidence identifies the host galaxy of \grb\ as the most
likely counterpart of \smm. The argument is supported on probabilistic
grounds and more importantly, the tentative detection of the host at
centimeter wavelengths (\S{\ref{sec:properties}}). The optically
visible portion of the host galaxy of \grb\ is blue with
(F814W$-K)\simeq 2.1$. The rest-frame B-band luminosity of this galaxy
($z=1.477$) is $M_B=-18.6$ (or $M_{AB}(B)=-18.8$). We have used the
formalism of Lilly et al.~(1995)\nocite{lth+95}, and assume
C(B,I)=$-$0.2 for the color term of the K-correction, appropriate for
late-type galaxies. This is similar to the host galaxy of
GRB\ts{970228} (\cite{bdk00}) both of which are subluminous ($L\sim
0.1 L_*$) relative to $L_*$ galaxies at their respective redshifts.
However, note that neither the magnitude nor the luminosity of this
galaxy is exceptional compared to other GRB host galaxies
(\cite{mm98}, \cite{hf99}, \cite{sch00}). The $M_B$ luminosity
corresponds to a rest-frame blue luminosity $L_B\simeq\nu\times L_\nu
= 2\times 10^{9}$ L$_\odot$, which is small compared to the bolometric
luminosity of $L_{\rm{Bol}}=4\times{10}^{12}$ L$_\odot$ derived in
\S{\ref{sec:properties}}. Despite the prodigious star formation, the
distribution of dust in the host galaxy must consist of regions of
significant internal extinction while still allowing the escape of
some UV light.

\section{Discussion and Conclusions}\label{sec:conclusions}

Multi-wavelength studies of GRBs and their host galaxies have the
potential both to determine the fraction of star formation that is
optically obscured (\cite{wbbn98}, \cite{tot99}, \cite{bn00}) and to
elucidate the relation between the optical/UV and submillimeter
selected galaxies.  To the degree that GRBs trace massive star
formation in the early universe (\cite{bkd+99}, \cite{gtv+00},
\cite{pgg+00}, \cite{bkd00}, \cite{rei01}), then the extreme
luminosity and the dust penetrating ability of $\gamma$-rays can be
used to select high redshift galaxies in an unbiased way without
regard to their emission properties.  Surveys for optical/UV galaxies
are magnitude-limited while, in contrast, the sample of GRB host
galaxies have magnitudes $V\sim 22-28$ which span the complete range
of model predictions (\cite{hf99}).  Obtaining reliable galaxy
identifications is a ongoing problem for submillimeter sources but the
early detection of an afterglow ensures that both an accurate position
and (in many cases) a redshift will be known for the galaxy.


In this paper we have identified a persistent, steep spectrum
submillimeter source and concluded that it is a dusty, starburst
galaxy, most likely the host galaxy of \grb. Although centimeter and
submillimeter observations of GRB host galaxies have only just begun
(\cite{stv+99}, \cite{bkf01}, \cite{hlm+00}), they are revealing
aspects of the sample that were not apparent from more extensive
optical/NIR observations. The detection of significant submillimeter
and centimeter radiation suggests that at least some GRB host galaxies
are undergoing prodigious star formation ($>500 M_\odot$ yr$^{-1}$).
Such results are important as they are a direct link between GRBs and
episodes of massive star formation (Berger et al. 2001).

Future deep submillimeter and centimeter radio observations of GRB
host galaxies hold some promise for exploring these differences in
their optical/UV properties. The current sample is somewhat limited.
Centimeter observations have not been carried out to the requisite
level ({\it i.e.}, at $z$=1, 10 $\mu$Jy$\simeq 10 M_\odot$ yr$^{-1}$
at 1.4 GHz) and to do so for the entire sample would require the
sensitivity of the Expanded VLA (EVLA), currently under development.
Submillimeter and millimeter observations have been undertaken toward
10 well-localized GRBs (\cite{sfkm98}, \cite{bbg+98}, \cite{kfs+99},
\cite{stv+99}, \cite{fbg+00}, \cite{bsf+00}) but for the most part the
existing upper limits would not be sufficient to rule out galaxies as
bright as \smm. Ramirez-Ruiz et al (2001\nocite{rtb01}) have
calculated that 20\% of GRB host galaxies are detectable at 350 GHz
with the SCUBA array on JCMT, a number that is close to the current
detection rate.

\acknowledgements We thank all the observers on the JCMT whose
programs were displaced in order to enable these target-of-opportunity
observations.  We are grateful to E. Kreysa and the MPIfR bolometer
team for providing MAMBO, and to R. Zylka for the MOPSI software
package. Thanks also to the IRAM staff for allowing flexible
scheduling at the 30m. This work was supported in part by grants from
the NSF, NASA, and private foundations to SRK, SGD, FAH, and RS,
Fairchild Fellowships to RS and TJG, Hubble Fellowship to DER,
Millikan Fellowship to AD, and Hertz Fellowship to JSB. IRAM is
supported by INSU/CNRS (France), MPG (Germany) and IGN (Spain).
Research at the Owens Valley Radio Observatory is supported by the
National Science Foundation through grant number AST 96-13717. DAF
thanks Rob Ivison and C. Carilli for useful conversations and an
anonymous JCMT referee for the catchy title. This research has made
use of NASA's Astrophysics Data System Abstract Service.


\clearpage
\begin{deluxetable}{ccccrrl}
\tabcolsep0in\footnotesize
\tablewidth{\hsize}
\tablecaption{Submillimeter and Millimeter Observations of
  \grb\label{tab:radio}\tablenotemark{a}}
\tablehead {
\colhead {}            &
\colhead {Freq.}       &
\colhead {Start}  &
\colhead {UT Start-End}    &
\colhead {TOS}         &
\colhead {Flux$\pm$rms}        &
\colhead {Notes}         \\
\colhead {Telescope}   &
\colhead {(GHz)}       &
\colhead {Date}            &
\colhead {(H:M)}  &
\colhead {(s)}         &
\colhead {(mJy)}   &
\colhead { }   
}
\startdata
JCMT &   660 &  Feb. 24 & 16:17 - 17:56 &  3600  &  $<37.8~~(3\sigma)$ & $\tau=0.055$\nl
JCMT &   350 &  Feb. 22 & 13:03 - 18:39 & 10800  &    4.34 $\pm$ 1.09 & $\tau=0.11$\nl
JCMT &   350 &  Feb. 23 & 11:08 - 19:30 & 18900  &    3.36 $\pm$ 0.72 & $\tau=0.09$ \nl
JCMT &   350 &  Feb. 24 & 16:17 - 19:29 &  6300  &    4.04 $\pm$ 1.14 & $\tau=0.055$ \nl
JCMT &   350 &  Mar. 01 & 11:54 - 12:50 &  2070  &    1.14 $\pm$ 2.06 & $\tau=0.08$ \nl
JCMT &   350 &  Mar. 02 & 14:54 - 18:05 &  7200  &    0.63 $\pm$ 0.94 & $\tau=0.05$\nl
JCMT &   350 &  Mar. 12 & 16:18 - 19:07 &  5400  &    3.98 $\pm$ 1.25 & $\tau=0.05$\nl
\hline
IRAM &   250 &  Feb. 22 & 22:33 - 06:43 &  8850  &    1.32 $\pm$ 0.54 & $\tau=0.45-0.51$\nl  
IRAM &   250 &  Feb. 25 & 00:28 - 07:55 &  9140  &    1.09 $\pm$ 0.32 & $\tau=0.17-0.29$\nl 
IRAM &   250 &  Feb. 26 & 01:12 - 08:52 &  8160  &    1.00 $\pm$ 0.33 & $\tau=0.23-0.28$\nl
IRAM &   250 &  Mar. 13 & 07:55 - 08:09 &   660  &    0.91 $\pm$ 0.94 & $\tau=0.08$\nl
IRAM &   250 &  Mar. 14 & 05:31 - 06:00 &  1330  &    1.31 $\pm$ 0.61 & $\tau=0.12$\nl
IRAM &   250 &  Mar. 18 & 22:33 - 22:48 &   830  &    0.78 $\pm$ 1.10 & $\tau=0.17$\nl
IRAM &   250 &  May  05 & 22:57 - 03:18 &  2300  &    1.13 $\pm$ 0.48 & $\tau=0.14-0.16$\nl
\hline
OVRO &   221 &  Feb. 23 & 07:26 - 17:40 & 18000  &    1.15 $\pm$ 5.00 & $\tau=0.25$\nl
OVRO &  98.5 &  Feb. 22 & 14:31 - 19:44 &  8400  &    0.30 $\pm$ 2.00  & $\tau>0.25$\nl
OVRO &  98.5 &  Feb. 23 & 07:26 - 17:40 & 18000  & $-$0.30 $\pm$ 0.80 & $\tau=0.25$\nl
\enddata
\tablenotetext{a}{From left to right the columns are the telescope
  used, the sky frequency at which the observations were made, the
  starting date, the UT start and end time, the time on source in
  seconds, the peak flux density at the position of \grb\ and the rms
  noise in milliJansky, and the zenith atmospheric opacity at 220 GHz
  (250 GHz at IRAM).  See \S\ref{sec:robs} for more details on the
  data calibration.}
\end{deluxetable}

\clearpage
\begin{deluxetable}{lcccc}
\tabcolsep0in\footnotesize
\tablewidth{\hsize}
\tablecaption{Magnitudes and Colors of K-band selected galaxies within
the SCUBA beam\label{tab:host}}
\tablehead {
\colhead {Galaxy}      &
\colhead {Aperture (\arcsec)}       &
\colhead {F814W}  &
\colhead {$K^a$}    &
\colhead {F814W$-K$}  
}
\startdata
A    & 1.42 & 21.786$\pm$0.010 & 18.233$\pm$0.034 & 3.553$\pm$0.035 \nl
B    & 1.60 & 22.931$\pm$0.028 & 18.398$\pm$0.035 & 4.533$\pm$0.044 \nl
C    & 1.96 & 22.124$\pm$0.018 & 19.049$\pm$0.041 & 3.074$\pm$0.044 \nl
D    & 1.42 & 24.934$\pm$0.151 & 21.102$\pm$0.113 & 3.832$\pm$0.188 \nl
Host & 0.45 & 25.58$\pm$0.13   & 22.745$\pm$0.274\tablenotemark{b} & 2.834$\pm$0.303 \nl
\enddata
\tablenotetext{a}{The error estimates do not include a zeropoint
  uncertainty of 0.1 mag (see \S\ref{sec:oobs}).}
\tablenotetext{b}{The host galaxy of \grb\ in the $K$ band is
  contaminated to some extent by the afterglow. An estimate from broad
  band modeling is that the afterglow contributes approximately 50\%
  of the total flux at this epoch. There is no afterglow component in
  the F814W band, since the host and afterglow were fit independently.
  After correcting for the afterglow contamination we find that the
  host color is bluer, {\it i.e.} (F814W$-K)\simeq 2.1$.}
\end{deluxetable}

\clearpage

\clearpage

\clearpage


\end{document}